\documentclass{fmj2010}
\usepackage{graphicx}

\usepackage{url}

\def\fermilat{\textit{Fermi}/LAT}
\def\fermi{\textit{Fermi}}

\begin{document}
   \title{The TANAMI Program}

   \author{Roopesh Ojha\inst{1}\thanks{rojha@usno.navy.mil}
          \and
          Matthias Kadler\inst{2,3,4}
          \and
          Moritz B\"ock\inst{2}
          \and
          Faith Hungwe\inst{5,6}
          \and
          Cornelia M\"uller\inst{2}
          \and
          Joern Wilms\inst{2}
          \and
          Eduardo Ros\inst{7,8}
          \and 
          the TANAMI Team
          }

   \institute{NVI/United States Naval Observatory, 3450 Massachusetts Ave, NW, Washington, DC 20392-5420
         \and
            Dr. Karl Remeis-Sternwarte \& ECAP, Sternwartstrasse 7, 96049 Bamberg, Germany 
             \and
             CRESST/NASA Goddard Space Flight Center, Greenbelt, MD 20771, USA
             \and 
             USRA, 10211 Wincopin Circle, Suite 500 Columbia, MD 21044, USA
             \and
             Department of Physics \& Electronics, Rhodes University, PO Box 94, Grahamstown 6140, South Africa
             \and
             Hartebeesthoek Radio Astronomy Observatory, PO Box 443, Krugersdorp 1740, South Africa
             \and
             Department d'Astronomia i Astrofisica, Universitat de Val\'encia, E-46100 Burjassot, Spain
             \and
             MPIfR, Auf dem H\"ugel 69, 53121 Bonn, Germany
             }

   \abstract{
 The TANAMI (Tracking AGN with Austral Milliarcsecond Interferometry) program provides comprehensive VLBI monitoring of extragalactic gamma-ray sources south of declination -30 degrees. Operating at two radio frequencies (8 and 22 GHz), this program is a critical component of the joint quasi-simultaneous observations with the \fermi\ Gamma-ray Space Telescope and ground based observatories to discriminate between competing theoretical blazar emission models. We describe the TANAMI program and present early results on the 75 sources currently being monitored. }

   \maketitle
%
\begin{figure*}
	\centering
	\vspace{39pt}
	\includegraphics[width=1.0\textwidth]{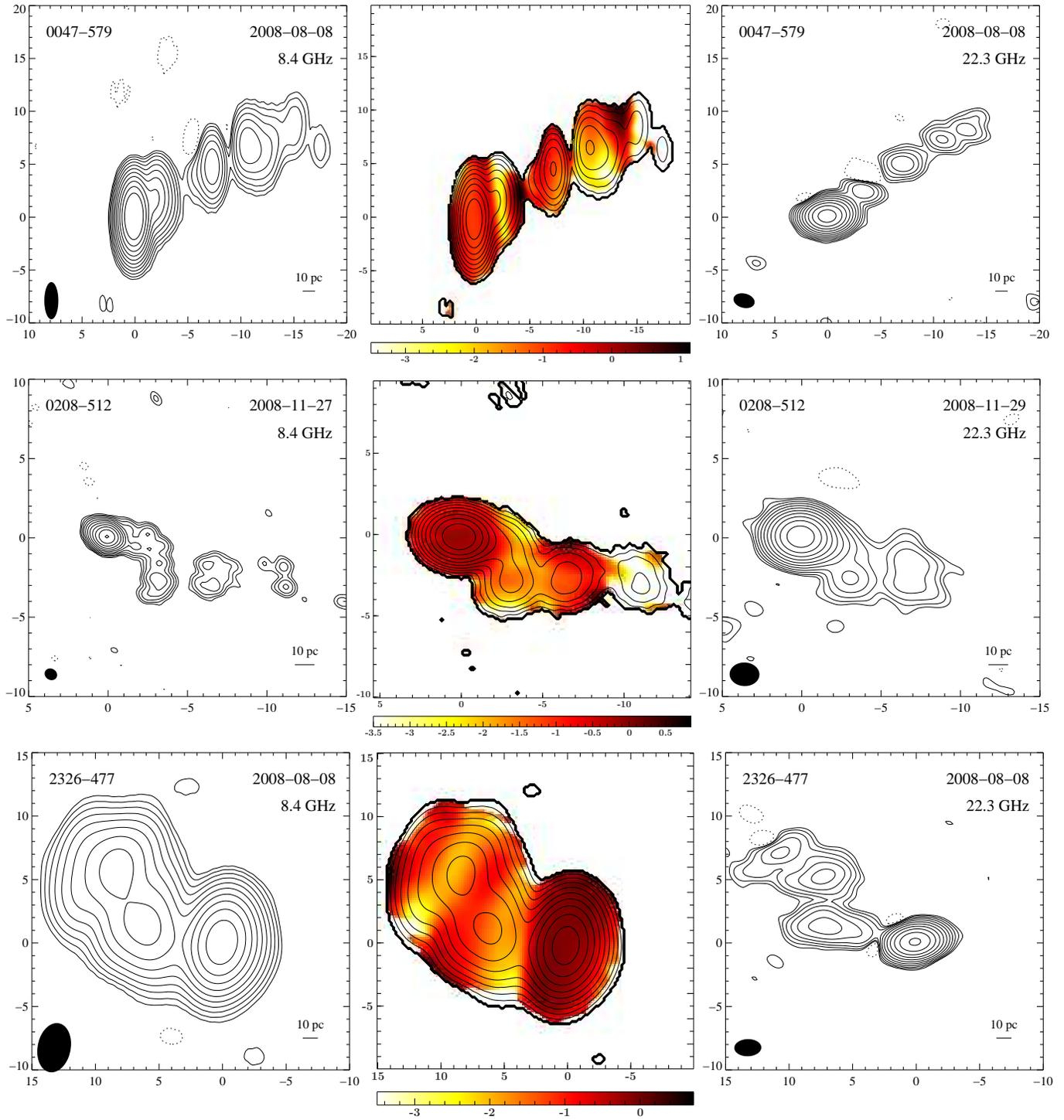}
   \caption{TANAMI images of four \fermi\ sources. Starting from the top, the four rows show images of 0047$-$579, 0208$-$512, 
   and 2326$-$477 respectively. In each row the left image shows the image at 8\, GHz and the image on the right at 
   22\,GHz at the same epoch. In the center of each row is the spectral index image made from the simultaneous images at 
   these two frequencies. Both the axes in all plots are labeled in milliarcseconds from the center of the image. The hatched 
   ellipse at the bottom left of each contour image represents the synthesized beam of the observing array. The color coding 
   in the spectral index image represents the spectral index defined as \mbox{$F_\nu \sim \nu^\alpha$}
            \label{fig:1}
           }
\end{figure*}

\section{Introduction}

Very Long Baseline Interferometry (VLBI) observations play a unique role in unraveling the physics of active galactic 
nuclei (AGN). They provide the only direct measurements of relativistic motion in AGN, thus measuring jet speeds, 
Doppler factors, opening and inclination angles of jets. With their unmatched resolution, VLBI observations can allow 
us to associate $\gamma$-ray flaring activity with structural changes on millarcsecond scales (such as jet-component 
ejections) helping to identify the location and extent of emission regions. 

VLBI observations have acquired particular salience in the age of \fermi.  Data from \fermilat\ in combination with other 
space and ground-based telescopes have made possible the quasi-simultaneous observations across the electromagnetic 
spectrum that have long been considered essential to distinguish between different models of AGN emission. 
The close connection between VLBI and \fermi\ observations is impressively demonstrated by the large number of VLBI-\fermi\ 
papers published and submitted in the past year, including many from the \fermilat\ collaboration, to which VLBI observations  
have contributed crucially needed data for the proper interpretation of $\gamma$-ray results. With VLBI data we have 
started to address some of the most crucial questions raised by the association of $\gamma$-ray emission 
with blazars.

\section{The TANAMI Program}

The indispensable role of parsec-scale monitoring of radio- and $\gamma$-ray bright AGN has lead to the establishment 
of a number of highly successful VLBI monitoring programs (see \cite{Lister2010} for a review) but all of these programs 
use northern hemisphere arrays that cannot observe much of the southern hemisphere. The TANAMI program is the only 
parsec scale monitoring program targeting AGN south of declination $-30^{\circ}$. Further, uniquely among comparable 
VLBI programs, TANAMI observations are made at two frequencies (8.4 and 22\,GHz). This lets us monitor the parsec-scale 
spectra of the cores and the brightest jet features, allowing us to contribute radio spectral indices of jet features to \fermi\ 
multiwavelength studies (e.g., \cite{Abdo2010a}, \cite{Chang2010}) besides probing emission processes along AGN 
jets (e.g., \cite{Mueller2010, Hungwe2010}). 

Since it covers that third of the sky not observed by other VLBI monitoring programs, TANAMI significantly improves the 
statistics for jet kinematics and flare-ejection studies. This region of the sky includes many interesting AGN (see below) 
and newly discovered $\gamma$-ray AGN can be followed up, often for the first time, with VLBI (e.g., \cite{Abdo2009a}). 
The TANAMI collaboration has also begun work with the ANTARES (\cite{Coyle2010}) and KM3NeT (\cite{Piattelli2010}) consortia, two neutrino telescopes 
that target the southern sky. \fermi\ $\gamma$-ray variability data and TANAMI-determined jet-ejection epochs will help 
develop data-filtering techniques to search for extragalactic neutrino point sources. This could usher us into an era of 
multimessenger astronomy. 

TANAMI observations are made using the telescopes of the Australian Long Baseline Array (LBA\footnote{The Long 
Baseline Array is part of the Australia Telescope which is funded by the Commonwealth of Australia for operation as a 
National Facility managed by CSIRO.}; e.g., \cite{Ojha2005}) and affiliated telescopes. TANAMI was able to significantly 
improve the $(u,v)$-coverage of the LBA by obtaining access to International VLBI Service (IVS) telescopes in Antarctica 
and Chile as well as Deep Space Network telescopes in Tidbinbilla, Australia. All telescopes that participate in TANAMI 
observations are listed, along with their diameters, in Table~1. At each epoch and each frequency, every source is 
typically observed for 6 scans of about 
10 minutes each. Typical $(u,v)$-coverage at both frequencies are shown in \cite{Mueller2010}. Our augmentation of 
the LBA has lead to the highest fidelity images for most of the sources observed by TANAMI. 

The initial sample of 44 TANAMI sources were selected based on previous (EGRET) $\gamma$-ray detection and/or 
radio flux density and luminosity. 
Under an MoU (Memorandum of Understanding) with the \textit{Fermi} collaboration TANAMI started monitoring observations of new \fermi\ sources 
through 2009 adding the new sources to our observing schedule while decreasing the observing cadence of sources 
showing limited radio-structural variability. The current TANAMI sample includes 75 sources of which 55 have been 
detected by \fermi. 53 TANAMI sources have 1FGL (\cite{Abdo2010c}) associations while 2 are tentative new detections 
(\cite{Boeck2010}).  
To date, 12 epochs (most at both frequencies) have been observed. Correlation, processing and imaging are 
progressing smoothly. Images and other results are available at our website\footnote{\url{http://pulsar.sternwarte.uni-erlangen.de/tanami/}} as soon as they are finalized.
For further details of the TANAMI program including details of calibration and imaging  see \cite{Ojha2010}.


   \begin{table}
      \caption[]{VLBI array for TANAMI observations.}
         \label{telescopes}
     $$ 
         \begin{array}{p{0.5\linewidth}l}
            \hline
            \noalign{\smallskip}
            Telescope     &  \mathrm{Diameter}  \\
            \noalign{\smallskip}
                                    &  \mathrm{(meters)}   \\            
            \noalign{\smallskip}
            \hline
            \noalign{\smallskip}
            Parkes, NSW, Australia & 64   \\
            Narrabri, NSW, Australia & 5\times22 \\
            Hobart, TAS, Australia & 26 \\
            Ceduna, SA, Australia & 30 \\
            Hartebeesthoek, S. Africa$^{\mathrm{a}}$ & 26 \\
            DSS43, ACT, Australia$^{\mathrm{b}}$ & 70 \\
            DSS45, ACT, Australia$^{\mathrm{b}}$ & 34 \\
            O'Higgins, Antarctica$^{\mathrm{c}}$ & 9 \\
            TIGO, Concepci\'on, Chile$^{\mathrm{c}}$ & 6 \\
            \noalign{\smallskip}
            \hline
         \end{array}
     $$ 
\begin{list}{}{}
\item[$^{\mathrm{a}}$] Not available since Sept 2008. Likely return Sept 2010.
\item[$^{\mathrm{b}}$] Operated by the Deep Space Network of the National Aeronautics and Space Administration, USA
\item[$^{\mathrm{c}}$] Operated by the German Bundesamt f\"ur Kartographie und Geod\"asie (BKG) \url{http://www.bkg.bund.de/nn_147094/EN/Home/homepage__node.html__nnn=true}
\end{list}
   \end{table}

\section{Results}

TANAMI is routinely producing VLBI images of high quality at 8 and 22\,GHz (X and K-band respectively). We 
show examples for three sources in Fig.~1. For each source we show the 8.4\,GHz image and the 22\,GHz image 
from the same epoch (on the left and right respectively). In the center of each row is shown the corresponding 
two-frequency spectral 
index image. It is important to note that the resolution of the lower frequency image is often {\it better} than that of the 
higher frequency image because the trans-oceanic telescopes in Antarctica and Chile cannot 
observe at 22\,GHz. 

These spectral index images were made by aligning the brightest pixels in the X and K-band images from the same epoch. 
The images at both frequencies have been convolved with the larger of the two beams of the individual images. The larger 
beam has also been used to produce the overlaid contours on these images. The color coding depicts 
the spectral index defined as \mbox{$F_\nu \sim \nu^\alpha$} i.e., a positive spectral indicates an inverted spectrum. 
Thus we are able to measure spectral indices of the cores and individual jet features and we are using these 
data to measure core shift, localize the central engine, calculate the opacity towards the central engine and identify the 
emission along the jet. In combination with data at other wavelengths we are modeling the SEDs of AGN. Note that 
the figures shown here are not corrected for coreshift.  

For a growing number of sources in our sample we have enough epochs of data to study their kinematics. We are 
fitting Gaussian components to jet features to track jet trajectories, measure their speeds, and derive their intrinsic 
parameters. When combined with SED modeling, these kinematic data address the relationship between the 
Doppler-boosting parameters for the radio and $\gamma$-ray emitting regions of the jets. 

   
   
   


TANAMI data have been and are being used in a number of studies that can broadly divided into two categories, 
individual source studies and statistical studies of the full sample or some subset, which are briefly described below. 

\subsection{Individual Source Studies}
Studies of individual TANAMI sources include:

\noindent$\bullet$ One of the first \fermilat\ publications addresses a bright $\gamma$-ray flare of the poorly studied 
source PKS\,1454-354 (\cite{Abdo2009a}). TANAMI contributed the first deep 8.4\,GHz VLBI image of this source revealing 
its core-jet structure. 

\noindent$\bullet$ TANAMI data on nine \fermilat\ sources were used to generate SEDs of the $\gamma$-ray selected 
LBAS blazars and investigate their broadband spectral propeties (\cite{Abdo2010a}).

\noindent$\bullet$ TANAMI data were used to construct the SED of PKS\,2052$-$47 during a LAT multiwavelength 
campaign (\cite{Chang2010}).

\noindent$\bullet$ TANAMI data are being used to study the highly variable BLLac  0537$-$441 which is one of the most 
luminous $\gamma$-ray blazars detected in the southern sky so far (\cite{Hungwe2010})

\noindent$\bullet$ TANAMI data were used to constrain the size of the $\gamma$-ray emitting region and for SED modeling 
of the nearest galaxy Centaurus A (\cite{Abdo2010b}). A multi-epoch, dual-frequency analysis of the innermost regions of this 
source is in progress (\cite{Mueller2010})

\subsection{First Epoch Results}
First epoch 8.4\,GHz results for the initial sample of 43 sources have been analyzed and presented in \cite{Ojha2010}.  
Using the classification scheme of \cite{Kellermann1998}, the initial sample has 33 single-side (SS) and 5 double-sided (DS)
sources with just one example each of the compact (C) and irregular (Irr) morphological types. Three sources do not have 
an optical identification. All of the quasars and BL Lacertae objects in the sample have an SS morphology
while all 5 DS sources are galaxies. The lone C source is optically unidentified while the only Irr source is a GPS 
galaxy 1718$-$649. 

The core and the total luminosity was calculated for all 38 initial
TANAMI sources that had published redshifts, assuming isotropic
emission. There is no significant difference in the distribution 
of luminosities of LBAS and non-LBAS sources. On the other hand, 
there is a clear relationship between luminosity and optical type 
with quasars dominating the high luminosity end of the distribution,
galaxies dominating the low luminosity end while the BL Lacertae 
objects fall in between.

The redshift distribution of the quasars and BLLacs in the TANAMI
sample is similar to those for the LBAS and EGRET blazars. There does
not appear to be any significant difference between the radio- and
$\gamma$-ray selected subsamples. The core brightness temperature ($T_\mathrm{B}$) 
limit of all initial TANAMI sources was calculated. The high end of the distribution of
calculated brightness temperatures is dominated by quasars and the low
end by BL Lacertae objects and galaxies. Of the 43 sources in the
sample, 14 have a maximum $T_\mathrm{B}$ below the equipartition value
of $10^{11}$~K (\cite{Readhead1994}), 30 below the inverse Compton limit
of $10^{12}$~K (\cite{Kellermann1969}), putting about a third of the
values above this limit. There is no significant difference in the brightness
temperature distribution of LBAS and non-LBAS sources.

A link between $\gamma$-ray 
emission and the parsec scale morphology of AGN has been sought (e.g., 
\cite{Taylor2007}). We fit circular Gaussians to the visibility data 
and measured the angle at which the innermost jet component appears
relative to the position of the core i.e. the opening angle. Of the
LAT AGN Bright Sample (LBAS) sources $78\%$ have an opening angle $>
30$ degrees while only $27\%$ of non-LBAS sources do. 
This result should be treated with great
caution as the sample size for this analysis is currently small 
but \cite{Pushkarev2009} report similar results.

If confirmed, the above result presents two possibilities: either the
LBAS jets have smaller Lorentz factors (since the width of the
relativistic beaming cone $\sim 1/\Gamma$) or LBAS jets are pointed
closer to the line of sight than $\gamma$-ray faint jets. The former scenario appears unlikely,
indeed the opposite effect is reported by \cite{Lister2009,
Kovalev2009}.

\section{Conclusions}

\fermi\ sources in the southern third of the sky are being monitored 
by the TANAMI program at about every two months. These high quality, dual frequency 
observations are producing spectral index images at milliarcsecond resolutions 
which are a crucial element in the multiwavelength study of AGN physics. For a 
subset of the TANAMI sample, the number of observed epochs is now sufficient 
for kinematic modeling to begin. When combined with jet-speed measurements, 
SED modeling across the electromagnetic spectrum will let us probe the relation 
between the Doppler-boosting parameters for the 
radio and $\gamma$-ray emitting regions of the jet. 

Studies of several individual AGN detected by \fermi\  have been enriched by data 
from the TANAMI program and multiwavelength analysis of a number of interesting sources 
are in progress. Statistical analysis of the growing TANAMI sample is 
providing broader insight into the tie between the low- and high-energy radiation from 
AGN.

\begin{acknowledgements}
We thank the \fermilat\ AGN group for the good collaboration.
This research has been partially funded by the Fermi Guest Investigator 
Program. 
This research has been partially funded by the Bundesministerium f\"ur
Wirtschaft und Technologie under Deutsches Zentrum f\"ur Luft- und
Raumfahrt grant number 50OR0808.  


\end{acknowledgements}

\end{document}